\documentclass[a4paper,fleqn,usenatbib]{mnras}
\DeclareMathAlphabet{\mathpzc}{OT1}{pzc}{m}{it}
\onecolumn
\usepackage{mathptmx}

\usepackage[T1]{fontenc}
\usepackage{ae,aecompl}

\usepackage{graphicx}	
\usepackage{amsmath}	
\usepackage{amssymb}	

\title[Evolution of spherical voids]{Evolution of density and velocity profiles 
of dark matter and dark energy in spherical voids}

\author[B. Novosyadlyj et al.]{
Bohdan Novosyadlyj$ $,\thanks{E-mail: bnovos@gmail.com}
Maksym Tsizh$ $ and Yurij Kulinich$ $
\\
$ $Ivan Franko National University of Lviv, Kyryla i Methodia str., 8, Lviv, 
79005, Ukraine}

\date{Accepted XXX. Received YYY; in original form ZZZ}

\pubyear{2016}

\begin{document}
\label{firstpage}
\pagerange{\pageref{firstpage}--\pageref{lastpage}}
\maketitle

\begin{abstract}
We analyse the evolution of cosmological perturbations which leads to the 
formation of large isolated voids in the Universe.
We assume that initial perturbations are spherical and all components of the 
Universe (radiation, matter and dark energy) are 
continuous media with perfect fluid energy--momentum tensors, which interact 
only gravitationally. Equations of the evolution of 
perturbations for every component in the comoving to cosmological background 
reference frame are obtained from equations of energy 
and momentum conservation and Einstein's ones and are integrated numerically. 
Initial conditions are set at the early stage of evolution 
in the radiation-dominated epoch, when the scale of perturbation is much larger 
than the particle horizon. Results show how the profiles 
of density and velocity of matter and dark energy are formed and how they 
depend 
on parameters of dark energy and initial conditions. 
In particular, it is shown that final matter density and velocity amplitudes 
change within range $\sim$4-7\% when the value of equation-of-state parameter 
of 
dark energy  $w$ vary in the range from --0.8 to --1.2, and change within 
$\sim$1\% only when the value of effective sound 
speed of dark energy vary over all allowable range of its values.
\end{abstract}



\section{Introduction}

The first voids as elements of the large-scale structure of the Universe have 
been revealed in 1978 by \cite{Joeveer78} and \cite{Gregory78} independently. 
The modern catalogues of voids \citep{Pan12,Sutter12,Nadathur14,Mao2016}, which 
have been created on the basis of the Sloan Digital Sky Survey (SDSS) data 
releases, contain the main data for about ten thousand of voids with radii 
spanning the range $\sim10-150$ Mpc located at redshifts up to $z\sim0.7$. The 
various aspects of their formation and distribution of sizes and amplitudes 
have 
been developed by \cite{Sheth2004} in the context of hierarchical scenarios of 
the large-scale structure formation. Investigations of last few years show that 
parameters of voids can be probe for cosmology and gravity theories since the 
void density profile, form, intrinsic dynamics, statistical properties and 
their 
dependences on voids sizes are sensitive to models of dark energy and modified 
gravity 
\citep{Biswas10,Bos12,Li12,Jennings13,Dai15,Li09,Clampitt13,Cai15,Hamaus15,
Falck16}. That is 
why the voids are extensively investigated 
using modern catalogues of galaxies as well as $N$-body simulations 
\citep{Gottloeber2003,Sutter14b,Sutter14c,Sutter14d,Wojtak2016,Demchenko16}.

The special interest for cosmology is the largest voids in the spatial 
distribution of galaxies, since they are more sensitive to the models of dark 
energy. Usually, it is assumed that dark energy is unperturbed in the voids or, 
at least, impact of its density perturbations on the peculiar motion and 
spatial 
distribution of galaxies is negligibly small. In this paper, we investigate the 
formation of isolated spherical voids as evolution of negative perturbations of 
density (underdense region) and velocity of matter together with dark energy 
ones from the early stage, when the scale of initial perturbation is much 
larger 
than particle horizon, up to current epoch. We analyse the influence of 
dynamical dark energy on such evolution and its dependence on initial 
conditions. We point attention to the evolution of density and velocity 
profiles 
of matter during void formation and its connection with universal density 
profile \citep{Hamaus14}. For this we have designed the code 
\citep{Novosyadlyj16} for integrating 
the system of equations obtained from the equations of relativistic 
hydrodynamics and gravitation
for description of evolution of spherical perturbations of densities and 
velocities in the three-component medium. The component `matter' consists
of dark matter ($\sim26$ per cent of total density) and typical baryonic matter 
($\sim4$ per cent), their dynamics at the large scales is well described by
the dust-like medium approach. 

\section{Model of spherical void: equations and initial conditions}

We assume that voids in spatial distribution of galaxies are formed as the 
result of the evolution of cosmological density perturbations 
in the three-component medium (radiation, matter and dark energy) with negative 
initial amplitudes and corresponding velocity profiles. It is believed that 
such 
perturbations are the result of quantum fluctuations of space--time metric in 
the inflationary epoch. They are randomly distributed in amplitude according to 
the normal law and are symmetrically relative to `$\pm$' for density and 
velocity perturbations at different spatial locations and different scales. We 
consider only scalar mode of perturbations, in which perturbations of density  
$\delta_N(t,r)$ and velocity $\mathpzc{v}_N(t,r)$ in every component $N$ are 
correlated because of survival of the growing solution  only at the stage when 
the scale of perturbation was larger than particle horizon. Positive 
perturbations lead to the formation of galaxies, galaxy clusters and so on, and 
the negative ones lead to the formation of voids. The first are well described 
by Press--Schechter formalism and theory of Gaussian peaks which are the base 
of 
halo theory of structure formation and its modern modifications based on the 
numerical $N$-body simulations (see \cite{Kulinich2013} and references 
therein). 
The spherical top-hat model of dust-like matter collapse is key for halo model. 
The spherical top-hat model of dark matter expansion in the $\Lambda$ cold dark 
matter ($\Lambda$CDM) model is important for semi-analytical studies of void 
evolution \citep{Sheth2004}. Here, we analyse the evolution of isolated 
negative 
density perturbations with spherical (but not top-hat!) initial profiles which 
form the voids with universal density profiles proposed by \cite{Hamaus14}.  

For that we use the system of seven differential equations in partial 
derivatives for seven unknown functions of two independent variables 
$\delta_{\rm m}(a,r)$, $\mathpzc{v}_{\rm m}(a,r)$, $\delta_{\rm de}(a,r)$, 
$\mathpzc{v}_{\rm de}(a,r)$, $\delta_{\rm r}(a,r)$, $\mathpzc{v}_{\rm r}(a,r)$ 
and $\nu(a,r)$ (see \cite{Novosyadlyj16} for details):  
\begin{eqnarray}
&&\dot{\delta}_{\rm m}-\frac{3}{2}(1+\delta_{\rm 
m})\dot{\nu}+\frac{1+\delta_{\rm m}}{a^2H}\left(\mathpzc{v}'_{\rm 
m}+\frac{2}{r}\mathpzc{v}_{\rm m}\right)+  
 \frac{\delta'_{\rm m}\mathpzc{v}_{\rm m}}{a^2H}=0,\label{m_cl0}\\
&&\dot{\mathpzc{v}}_{\rm m}+\frac{\mathpzc{v}_{\rm 
m}}{a}+\frac{\nu'}{2a^2H}+\frac{2\mathpzc{v}_{\rm 
m}}{a^2H}\left(\mathpzc{v}'_{\rm m}+\frac{\mathpzc{v}_{\rm m}}{r}\right)+
\frac{\dot{\delta}_{\rm m}\mathpzc{v}_{\rm m}}{1+\delta_{\rm 
m}}=0,\label{m_cl1}\\
&&\dot{\delta}_{\rm de}+\frac{3}{a}(c_{\rm s}^2-w)\delta_{\rm 
de}+(1+w)\left[\frac{\mathpzc{v}'_{\rm de}}{a^2H}+\frac{2\mathpzc{v}_{\rm 
de}}{a^2Hr}
-9H(c_{\rm s}^2-w)\int{\mathpzc{v}_{\rm de}dr}-\frac{3}{2}\dot{\nu}\right] 
\nonumber \\
&&\hskip7cm+(1+c_{\rm s}^2)\left[\frac{\delta'_{\rm de}\mathpzc{v}_{\rm 
de}}{a^2H} +\frac{\delta_{\rm de}}{a^2H}\left(\mathpzc{v}'_{\rm de}+
\frac{2}{r}\mathpzc{v}_{\rm de}\right)-\frac{3}{2}\delta_{\rm 
de}\dot{\nu}\right]=0,\label{de_cl0}\\ 
&&\dot{\mathpzc{v}}_{\rm de}+(1-3c_{\rm s}^2)\frac{\mathpzc{v}_{\rm 
de}}{a}+\frac{c_{\rm s}^2\delta'_{\rm de}}{a^2H(1+w)}+ 
\left(1+\frac{1+c_{\rm s}^2}{1+w}\delta_{\rm de}\right)\frac{2\mathpzc{v}_{\rm 
de}}{a^2H} \left(\mathpzc{v}_{\rm de}'+\frac{\mathpzc{v}_{\rm 
de}}{r}\right)\nonumber\\
&&\hskip5cm+\frac{\nu'}{2a^2H}+\frac{1+c_{\rm s}^2}{1+w}\left[\dot{\delta}_{\rm 
de}\mathpzc{v}_{\rm de}+\delta_{\rm de}\dot{\mathpzc{v}}_{\rm 
de}+(1-3w)\frac{\delta_{\rm de}}{a}\mathpzc{v}_{\rm de}+\frac{\nu'\delta_{\rm 
de}}{2a^2H}\right]=0, \label{de_cl1}\\  
&&\dot{\delta}_{\rm r}-2(1+\delta_{\rm 
r})\dot{\nu}+\frac{4}{3}\frac{1+\delta_{\rm r}}{a^2H}\left(\mathpzc{v}'_{\rm 
r}+\frac{2}{r}\mathpzc{v}_{\rm r}\right)+  
 \frac{4}{3}\frac{\delta'_{\rm r}\mathpzc{v}_{\rm r}}{a^2H}=0.\label{rel_cl0}\\ 
&&\dot{\mathpzc{v}}_{\rm r}+\frac{\nu'}{2a^2H}+\frac{\delta'_{\rm 
r}}{4a^2H(1+\delta_{\rm r})}+
\frac{\dot{\delta}_{\rm r}\mathpzc{v}_{\rm r}}{1+\delta_{\rm 
r}}=0,\label{rel_cl1}\\
&&\frac{1-\nu}{3a^2}\left(\nu''+\frac{2}{r}\nu'\right)-H^2(a\dot\nu+\nu)=
\frac{H^2_0}{a^3}\left(\Omega_{\rm m}\delta_{\rm m}+
\Omega_{\rm r}a^{-1}\delta_{\rm r}+\Omega_{\rm de}a^{-3w}\delta_{\rm 
de}\right). 
\label{ee00_l2}
\end{eqnarray}
Here $\Omega$-s denote the mean densities of the components in the unit of the 
critical one at the current epoch, $w\equiv p_{\rm de}/\rho_{\rm de}$ is the 
equation-of-state parameter of dark energy, $c_{\rm s}$ is the effective speed 
of sound of dark energy in its proper frame,  $H(a)\equiv d\ln{a}/dt$ is the 
Hubble parameter, which defines the rate of the expansion of the Universe and 
is 
a known function of time for given cosmology and the model of dark energy, 
$$H(a)=H_0\sqrt{\Omega_{\rm r}a^{-4}+\Omega_{\rm m}a^{-3}+\Omega_{\rm 
de}a^{-3(1+w)}},$$ 
were $H_0$ is the Hubble constant. The independent variables of the system of 
equations (\ref{m_cl0})-(\ref{ee00_l2}) are scale factor $a$ and radial 
comoving 
coordinate $r$, over-dot and prime denote the derivatives with respect to them. 
It is assumed that unperturbed space--time, cosmological background, is 
Friedman--Robertson--Walker one with zero spatial curvature. In the region, of 
void it is perturbed so that metric there is as follows:
\begin{equation}
ds^2=e^{\nu(t,r)}dt^2-a^2(t)e^{-\nu(t,r)}[dr^2+r^2(d\theta^2+\sin^2\theta 
d\varphi^2)] \label{ds_sph},
\end{equation}
where the metric function $\nu(t,r)$ defines the local deviation of curvature 
of 
3-space from zero. At the late stages, when the scale of perturbation is much 
smaller than the particle horizon, it is the doubled gravitational potential in 
the Newtonian approximation of equation (\ref{ee00_l2}). The density and 
3-velocity perturbations $\delta_N$  and $\mathpzc{v}_N$ are defined in 
coordinates, which are comoving to the  unperturbed cosmological background 
(see 
subsection 2.2 in \cite{Novosyadlyj16}), Newtonian gauge. Thus, the velocity 
perturbation coincides with definition of peculiar velocity of galaxies (see, 
for example, \cite{Peebles80}).

To take into account the Silk damping effect for radiation, we have added into 
equations (\ref{rel_cl0}) and (\ref{rel_cl1}) the terms $\delta_{\rm 
r}k_D/H/a^2$ and $\mathpzc{v}_{\rm r}k_D/H/a^2$ accordingly, where the scale of 
damping $k_D$ was computed by formula 10 from \cite{Hu95}.

To solve the system of equations (\ref{m_cl0})--(\ref{ee00_l2}), the initial 
conditions must be set. Let us relate the initial amplitude of given 
perturbation 
with mean-square one given by the power spectrum of cosmological perturbations. 
For this, we define the initial conditions in the early Universe, when 
$\rho_{\rm r}\gg\rho_{\rm m}\gg\rho_{\rm de}$, the physical size of the 
perturbation is much larger than the particle horizon ($k^{-1}\gg ct$, where 
$k^{-1}$ 
is a linear scale of perturbation in comoving coordinates) and amplitudes are 
small ($\delta,\,\mathpzc{v},\,\nu\,\ll\,1$). Without the loss of generality, the unknown 
functions can be presented in the form of separated variables
$$\nu(a,r)=\tilde{\nu}(a)f(r), \quad \delta_{N}(a,r)=\tilde{\delta}_{N}(a)f(r),  
\quad \mathpzc{v}_{N}(a,r)=\tilde{\mathpzc{v}}_{N}(a)f'(r),$$
where $f(0)=1$ and $f'(r)\propto r$ near the centre $r=0$. Ordinary 
differential 
equations for amplitudes $\tilde{\nu}(a)$, $\tilde{\delta}_{N}(a)$, 
$\tilde{\mathpzc{v}}_{N}(a)$ are obtained from general system of equations 
(\ref{m_cl0})-(\ref{ee00_l2}) by their expansion in Taylor series near the 
centre. The analytical solutions of equations for the amplitudes for the 
radiation-dominated epoch (matter and dark energy can be treated as test 
components) in  the `superhorizon' asymptotic give the simple relation for 
them: 
\begin{equation}
\tilde{\delta}^{\rm init}_{\rm r}=\frac{4}{3}\tilde{\delta}_{\rm m}^{\rm 
init}=\frac{4}{3(1+w)}\tilde{\delta}_{\rm de}^{\rm init}=-\tilde{\nu}^{\rm 
init}=C, \quad \tilde{\mathpzc{v}}^{\rm init}_{\rm r}=\tilde{\mathpzc{v}}_{\rm 
m}^{\rm init}=\tilde{\mathpzc{v}}_{\rm de}^{\rm init}=\frac{C}{4a_{\rm 
init}H(a_{\rm init})},\label{init1}
\end{equation}
where $C$ is the integration constant, value of which defines the initial 
amplitudes of perturbations in all components. 

We set the value of $C$ in the units of mean-square amplitude of cosmological 
perturbations, which is defined from modern observations. The \textit{Planck} + 
\textit{HST} + WiggleZ + SNLS3 data tell that amplitude $A_{\rm s}$ and 
spectral 
index $n_{\rm s}$ of power spectrum of initial perturbations of curvature 
$\mathcal{P_{\rm r}}(k)=A_{\rm s}(k/0.05)^{n_{\rm s}-1}$ are the following: 
$A_{\rm s}=2.224\times 10^{-9},\,\,n_{\rm s}=0.963$ \citep{Sergijenko15}. Since 
for perturbations with $ak^{-1}\gg ct$, the power spectrum perturbations of 
curvature $\mathcal{P_{\rm 
r}}\equiv\frac{9}{16}\left\langle\nu\cdot\nu\right\rangle$ is constant in time 
in the matter-dominated and radiation-dominated epochs, in the range of scales 
$0.01\le k \le0.1$ Mpc$^{-1}$ the initial amplitude which correspondents to 
mean-square one is $\sigma\approx\frac{4}{3}\sqrt{A_{\rm 
s}}\left(\frac{0.05}{k}\right)^{\frac{n_{\rm 
s}-1}{2}}\approx(6.1-6.7)\times10^{-5}$. In the computations we set 
$C=-1\times10^{-4}$ and $-2\times10^{-4}$ at $a_{\rm init}=10^{-6}$ for 
$k=0.05$ 
Mpc$^{-1}$, that corresponds  $\approx1.6$ and $3.2\sigma$ accordingly.

\begin{figure*}
\includegraphics[width=0.33\textwidth]{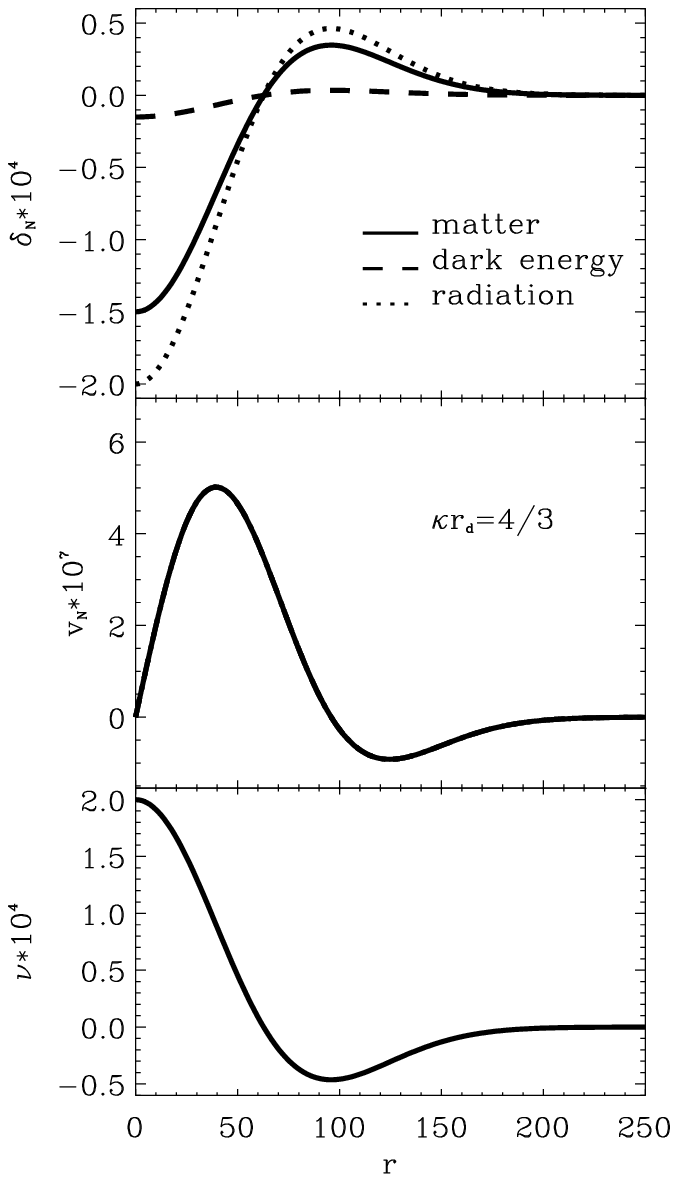} 
\includegraphics[width=0.33\textwidth]{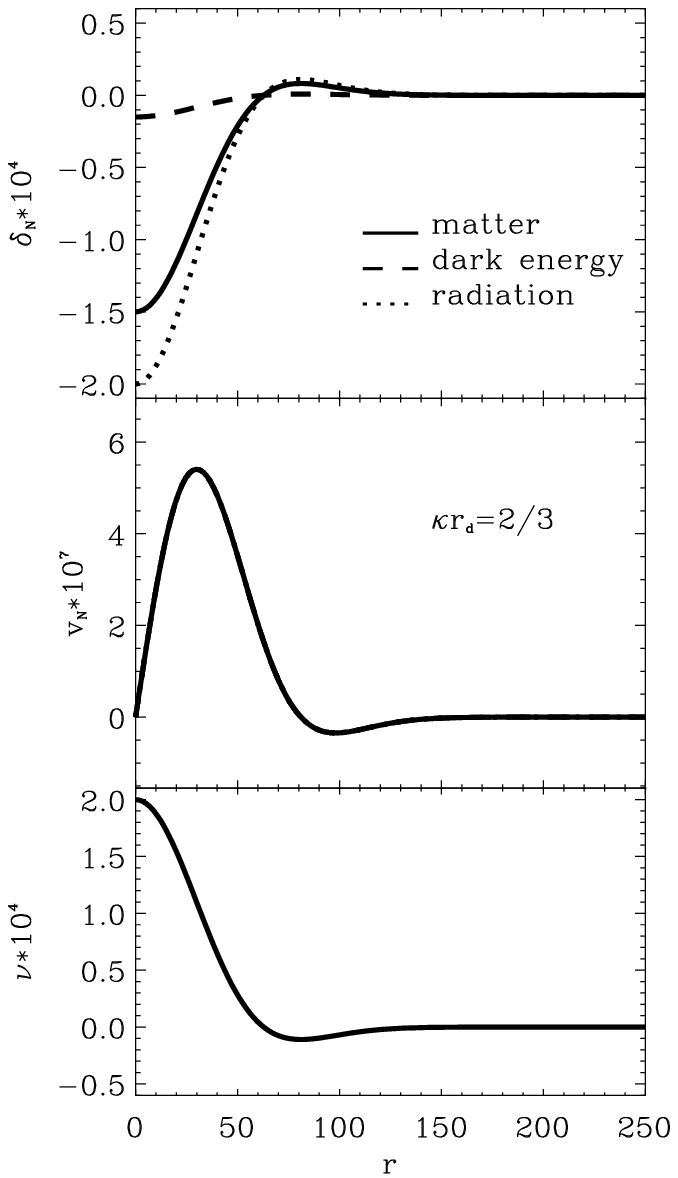}
\includegraphics[width=0.33\textwidth]{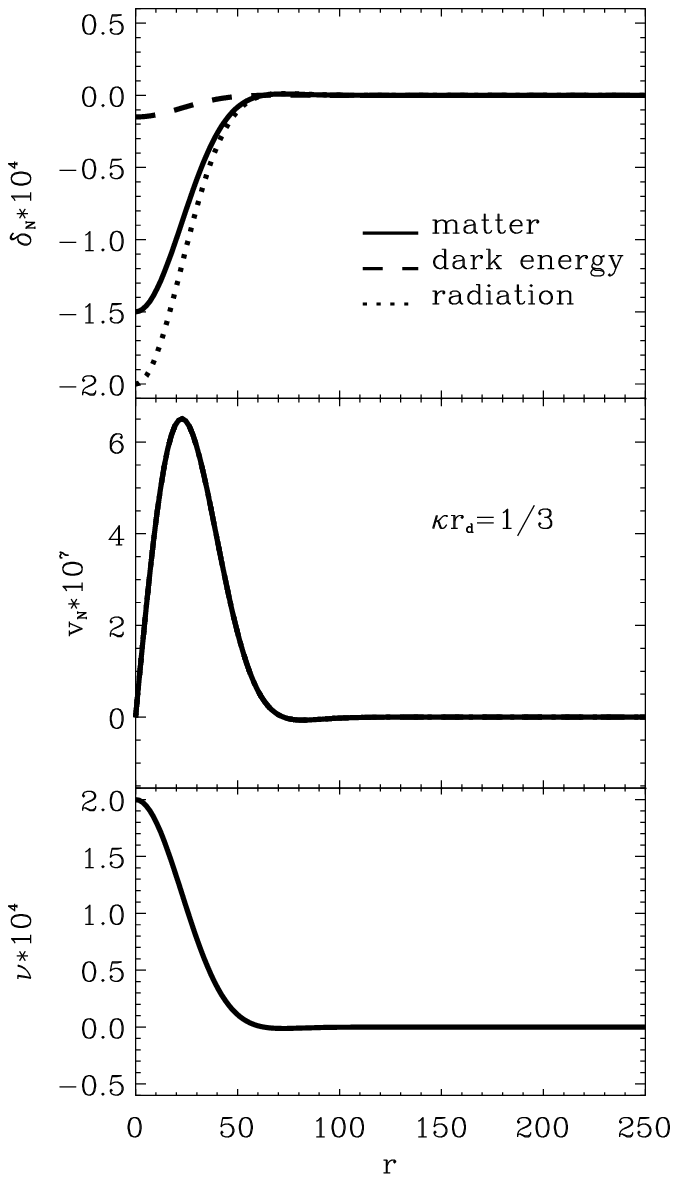}  
\caption{The initial profiles of density (top panels) and velocity (central 
panels) perturbations of dark matter (solid line), dark energy (dashed line) 
and 
relativistic components (dotted line). In the bottom panels the corresponding 
initial profiles of metric function $\nu(a_{\rm init},r)$ are presented. 
The parameters of initial profile are $\kappa^{-1}=62.8$ Mpc, $r_{\rm 
d}\approx83.7$ Mpc (left-hand column), $r_{\rm d}\approx41.9$ Mpc (middle 
column) and 
$r_{\rm d}\approx20.9$ Mpc (right-hand column). Other input parameters are: 
$H_0=70$ km/s/Mpc, $\Omega_{\rm m}=0.3$, $\Omega_{\rm de}=0.7$, $w=-0.9$, 
$c_{\rm s}=0$, $C=-2\times10^{-4}$. }  
\label{init}
\end{figure*}

In this work, we study the formation of the spherical voids with initial 
profile 
of density perturbations 
$f(r)=(1-\kappa^2 r^2)e^{-r^2/r_{\rm d}^2}$, where $\kappa$ defines the size of 
the void $r_{\delta=0}=\kappa^{-1}$, where the density perturbation becomes 
zero, $\delta_{\rm m}(r_{\delta=0})=0$. The scale of decaying $r_{\rm d}$ 
defines the parameters (position and amplitude) of overdense shell, which 
surrounds the void. Really, the second local extremum of $f(r)$ is at $r_{\rm 
sh}=\sqrt{r_{\rm d}^2+\kappa^{-2}}$, where $\delta_{\rm sh}\equiv 
max\{\delta_{\rm m}\}=-\kappa^2r_{\rm d}^2Ce^{-1-1/\kappa^2r_{\rm d}^2}$. In 
this paper we analyse the perturbations with initial $\kappa^2r_{\rm 
d}^2=4/3,\,2/3,\,1/3$. In Fig. \ref{init}, we show the initial profiles of 
$\delta_{\rm m}(a_{\rm init},r)$, $\mathpzc{v}_{\rm m}(a_{\rm init},r)$, 
$\delta_{\rm de}(a_{\rm init},r)$, $\mathpzc{v}_{\rm de}(a_{\rm init},r)$, 
$\delta_{\rm r}(a_{\rm init},r)$, $\mathpzc{v}_{\rm r}(a_{\rm init},r)$ and 
$\nu(a_{\rm init},r)$ for $r_{\delta=0}=62.8$ Mpc, $C=-2\times10^{-4}$ and
three values of  $\kappa^2r_{\rm d}^2$. They have the same size $r_{\delta=0}$ 
and the central magnitude of $\delta_{N}(a_{\rm init},0)$ and $\nu(a_{\rm 
init},0)$ but different position and amplitude of overdensity shell. 
The ratio of amplitudes is $\delta_{\rm sh}^{(4/3)}:\delta_{\rm 
sh}^{(2/3)}:\delta_{\rm sh}^{(1/3)}\approx 38:9:1$ accordingly to 
$\kappa^2r_{\rm d}^2=4/3,\,2/3,\,1/3$. In the units of magnitudes of central 
density perturbations, the magnitudes of overdense shells 
for these profiles are 0.23, 0.055 and 0.006.

\section{Method of numerical integration and parameters of models}

For numerical integration of the system of equations 
(\ref{m_cl0})-(\ref{ee00_l2}) with initial conditions (\ref{init1}), we have
designed the computer code 
\textit{npdes.f}\footnote{$\mathrm{http://194.44.198.6/\sim 
novos/npdes.tar.gz}$}, which 
implements the modified Euler method taking into account the derivatives from 
the forthcoming step and improving the results by 
iterations  \citep{Leveque1998}. This scheme of integration is resistant to the 
numerical spurious oscillations, and is fast and 
precise enough. For example, the Hamming method of prediction and correction of 
fourth-order of precision with five iterations at each step 
need three times more processor time for the same precision of final result. 
The 
step of integration was posed as variable: $\Delta a=a/N_a$, 
where number $N_a$ was picked up so that the numerical precision of the result 
of integration at $a=1$ was not worse than 0.1\%. 
In all computations presented here we assumed $N_a=3\times10^{6}$.   

The numerical derivatives with respect to $r$ in the grid with constant step 
$\Delta r=R_{\rm m}/N_{\rm r}$, where $R_{\rm m}$ is radius of spatial region 
of 
integration, were evaluated with help of third order polynomial by method of 
Savitzky--Golay convolution \citep{Savitzky64}: 
$y'_i=[3(y_{i+1}-y_{i-1})/4-(y_{i+2}-y_{i-2})/12]/\Delta r$. The computation 
experiments have shown that optimal values of space grid
parameters are $R_{\rm m}\sim8r_{\delta=0}$ and $N_{\rm r}=300$. 

In the test computations we have found that the spurious oscillations with 
growing amplitude appear in the dark energy, when its effective sound speed 
$c_{\rm s}>0.01c$, which was expected for elastic component 
\citep{Leveque1998}. 
To remove them we used the Savitzky--Golay convolution filter 
\citep{Savitzky64} 
with parameters $n_l=12,\,n_{\rm r}=12,\,m=6$, by which  the space-dependences 
of derivatives $\dot{\delta}_{\rm de}$ and $\dot{\mathpzc{v}}_{\rm de}$ were 
smoothed at each step of integration by $a$. Such smoothing practically does 
not 
influence on the final result of integration, which is confirmed by comparison 
of the results with smoothing and without it for case of the dark energy model 
with $c_{\rm s}=0$, for which spurious oscillations do not appear. 
The maximum difference is less than 4 per cent for density perturbation and 1 
per cent for velocity perturbation of dark energy in the region of maximum 
amplitude of velocity perturbation.

The computer code \textit{npdes.f} has been tested by comparison of the results 
of the integration by code with
(1) known analytical solutions for density and velocity perturbations in 
conformal-Newtonian gauge for radiation-dominated and matter-dominated 
Universes, (2) results of integration of linear perturbation by code 
\textit{camb.f} \citep{camb} and (3) results of integration by code 
\textit{dedmhalo.f} \citep{Novosyadlyj16}, developed on the basis of 
\textit{dverk.f} \footnote{$\mathrm{http://www.cs.toronto.edu/NA/dverk.f.gz}$} 
for the amplitude of the spherical perturbations in the centre of halo. 
In all the cases, deviations did not exceed a few tenths of a per cent, 
which means that precision of the integration is high enough for our studies.

The input parameters of the program are: the Hubble parameter $H_0$, the 
density 
parameters of all components $\Omega_{\rm r}$, $\Omega_{\rm de}$, $\Omega_{\rm 
m}=1-\Omega_{\rm de}-\Omega_{\rm r}$, the equation-of-state parameter of dark 
energy $w$, the speed of sound of dark energy $c_{\rm s}$, the initial 
amplitude 
of perturbation $C$, the parameters of profile $f(r)$ of initial perturbation 
$\kappa$ and $r_{\rm d}$. 

The value of density parameter of dark energy $\Omega_{\rm de}$ is determined 
on 
the basis of current observational data with accuracy $\sim2$ per cent and its 
mean 
value is close to 0.7 \citep{WMAP9,Planck2015,Sergijenko15}. The 
equation-of-state parameter of dark energy is determined worse, with accuracy 
$\sim$5--7 per 
cent and its mean value is close to --1. The value of effective speed of sound 
of dark energy is not constrained by cosmological observational data which are 
obtained up to now (see, for example, \cite{Sergijenko15} and references 
therein). For the estimation of the sensitivity of void parameters to $w$ and 
$c_{\rm s}^2$ we analyse the formation of voids in the cosmological models with 
dark energy with $\Omega_{\rm de}=0.7$, $w\in[-0.8,\,-1.2]$ and 
$c_{\rm s}\in[0,\,1]$. Other cosmological parameters in computations are fixed: 
$\Omega_{\rm r}=4.17\times10^{-5}$, $\Omega_{\rm m}=0.3-\Omega_{\rm r}$, 
$H_0=70$ km s$^{-1}$ Mpc$^{-1}$.

The calculations were performed for the different values of the parameters 
$\kappa$, $r_{\rm d}$, $C$, $w$ and $c_{\rm s}$. Each run with a fixed set of 
parameters produce the main output file which contain for each $a_i=a_{\rm 
init}\times10^{3i/20}$ with interval $\Delta a_i=0.29a_i$ ($i=0,\,...,\,40$) 
profiles $\delta_{\rm de}(a_i,r_j)$, $\mathpzc{v}_{\rm de}(a_i,r_j)$, 
$\nu(a_i,r_j)$, $\delta_{\rm m}(a_i,r_j)$, $\mathpzc{v}_{\rm m}(a_i,r_j)$, 
$\delta_{\rm r}(a_i,r_j)$, $\mathpzc{v}_{\rm r}(a_i,r_j)$ ($j=1,\,...,\,300$). 
The example of output file can be found in the package with code.

\begin{figure*}
\includegraphics[width=0.33\textwidth]{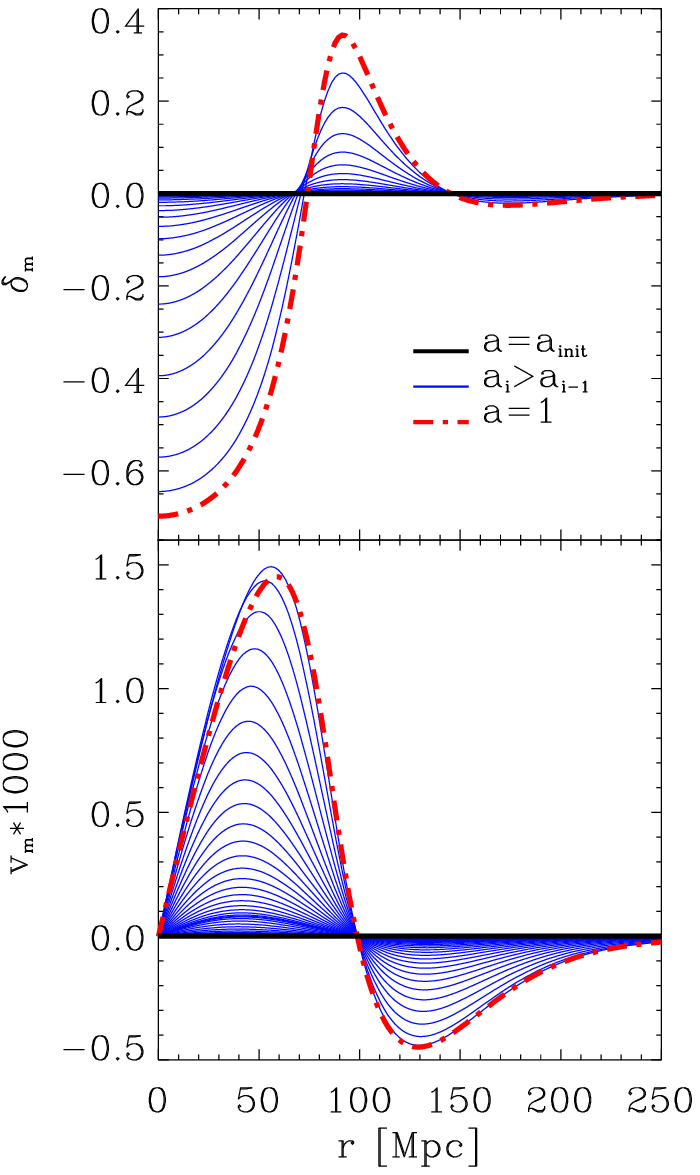} 
\includegraphics[width=0.33\textwidth]{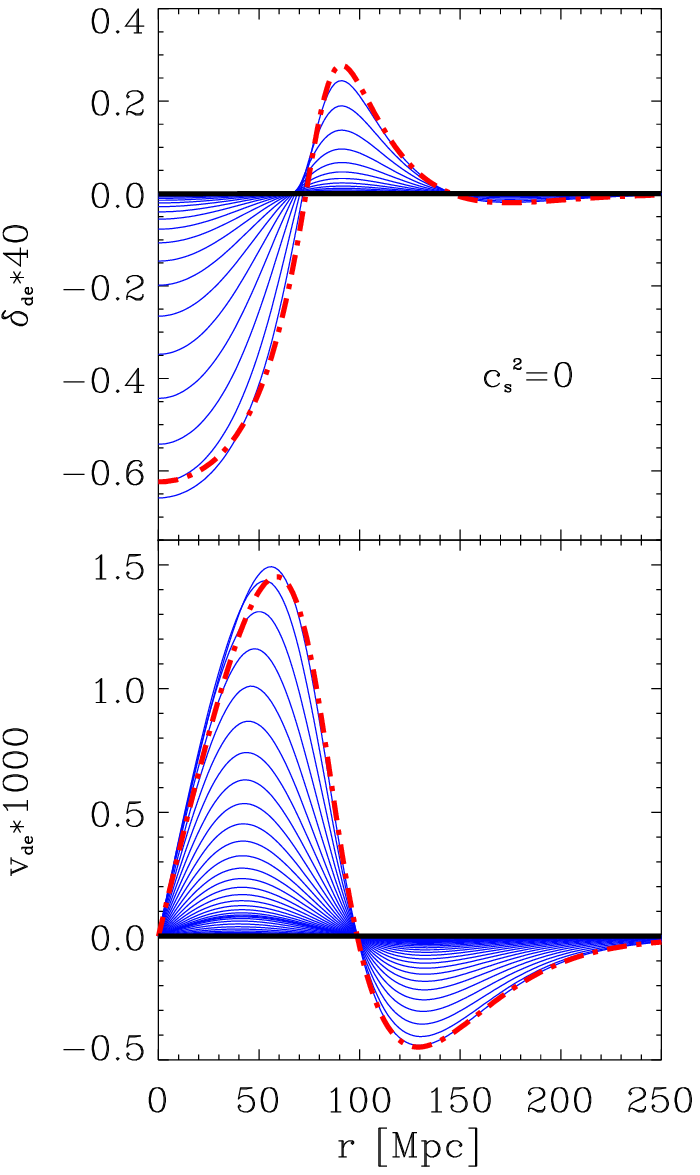}
\includegraphics[width=0.33\textwidth]{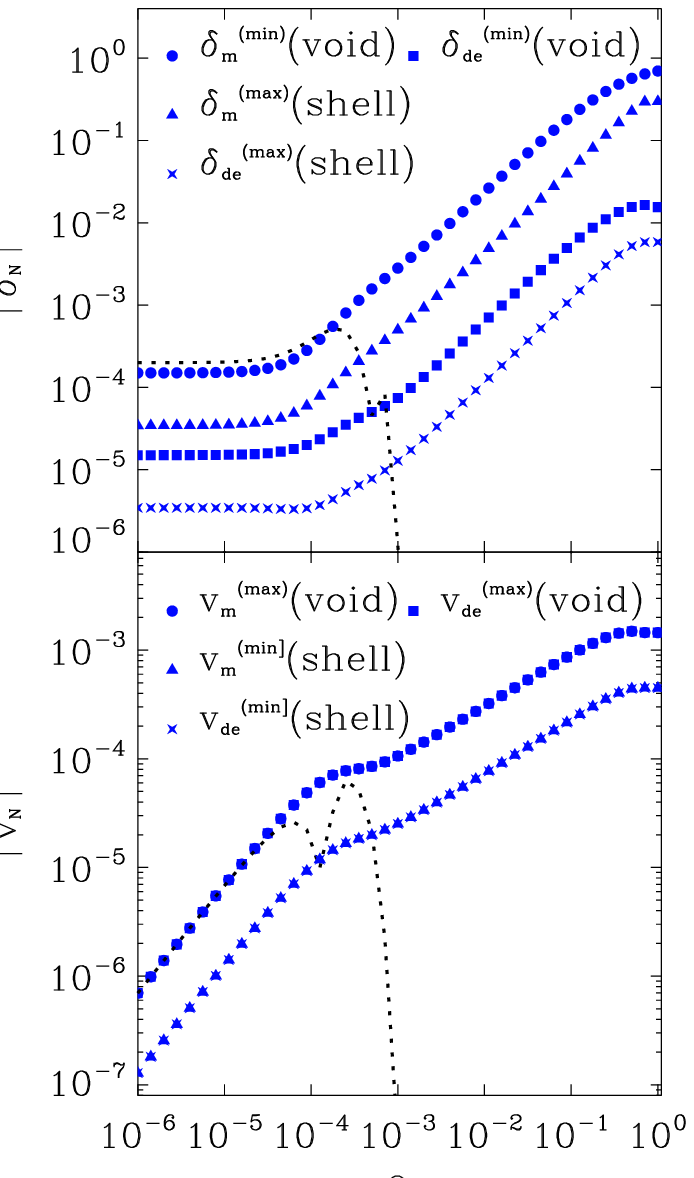}
\caption{Formation of void with initial profile parameters $\kappa^{-1}=62.8$ 
Mpc and $\kappa r_{\rm d}=4/3$ in dark matter (left-hand column) and dark 
energy 
with $c_{\rm s}^2=0$ (central column). In the right-hand column the evolution 
of 
absolute values of amplitudes of density (top panel) and velocity (bottom 
panel) 
perturbations for void and overdense shell are shown. Other cosmological and 
dark energy parameters are the same as in Fig. \ref{init}. The dotted lines 
show 
the evolution of absolute values of amplitudes of density and velocity 
perturbations for radiation component in the void only.}   
\label{cs00}
\end{figure*}

\begin{figure*}
\includegraphics[width=0.33\textwidth]{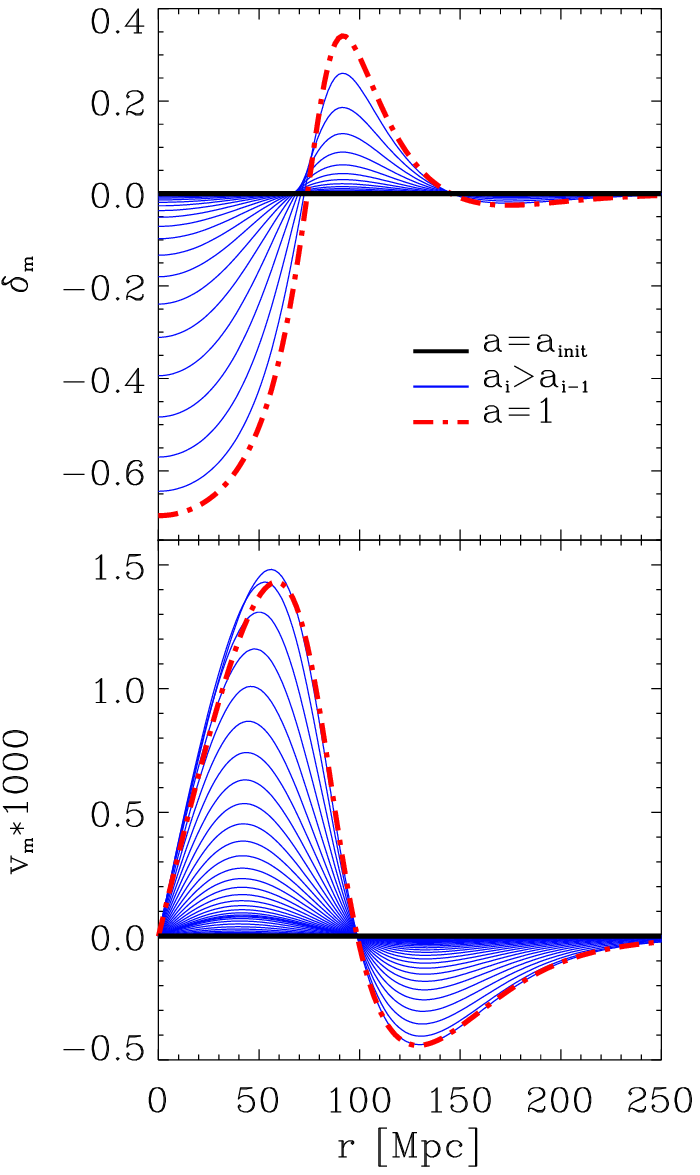} 
\includegraphics[width=0.33\textwidth]{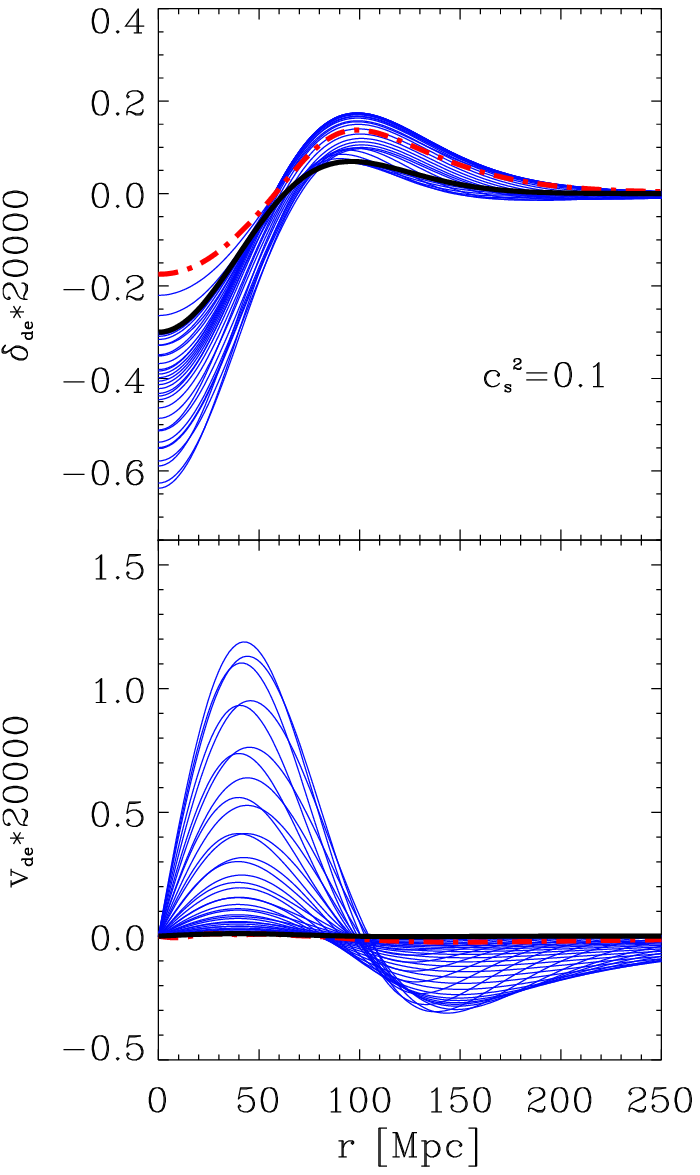}
\includegraphics[width=0.33\textwidth]{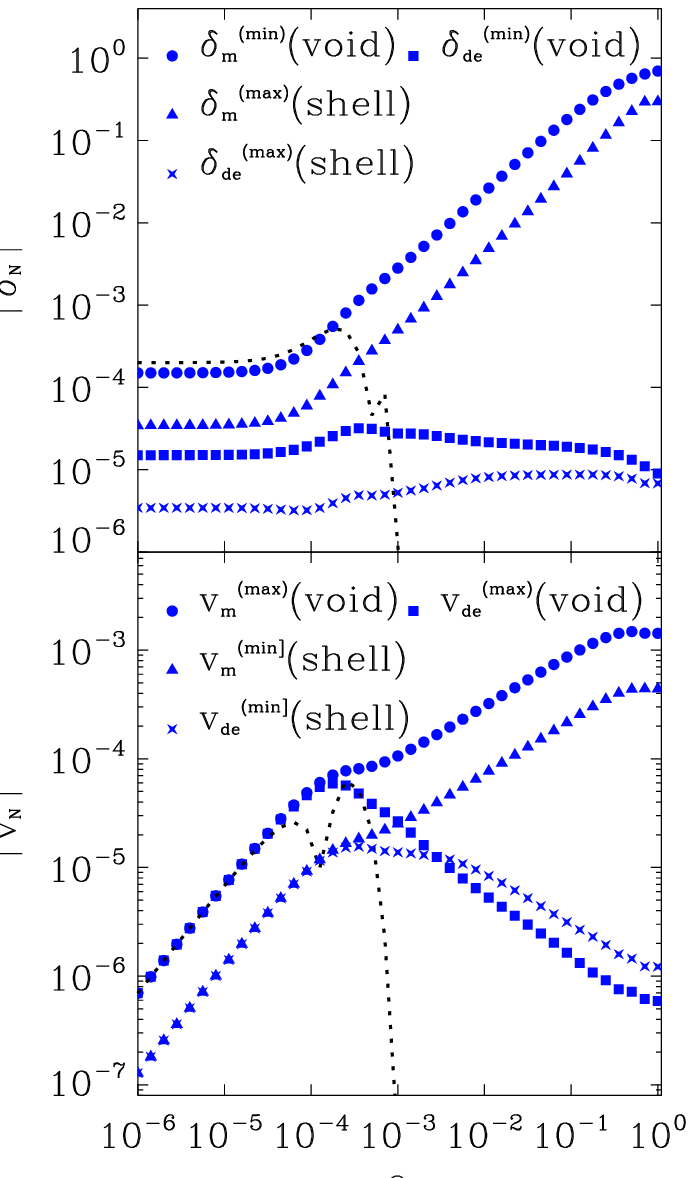}  
\caption{The same as in Fig. \ref{cs00} for dark energy with $c_{\rm s}^2=0.1$.}
\label{cs01}
\end{figure*}

The step-by-step guide of the performed computations is as follows. (i) To 
analyse the void formation in the models with dark 
energy of different perturbation ability we run the code for $c_{\rm 
s}^2=0,\,0.001,\,0.01,\,0.1,\,0.2,\,0.5,\,1$ and fixed 
$w=-0.9$,$\kappa^{-1}=r_{\delta=0}=62.8$ Mpc, $\kappa^2r_{\rm d}^2=4/3$, 
$C=-2\times10^{-4}$. (ii) To analyse the void formation in the
models with dark energy of different elastic ability we run the code for 
$w=-0.8,\,-0.9,\,-1,\,-1.1,\,-1.2$, fixed $c_{\rm s}^2=0.1$ 
and the same $\kappa$, $r_{\rm d}$ and $C$. (iii) To analyse the dependence of 
formation of overdensity shells on initial profile we run the code for
$\kappa^2r_{\rm d}^2=4/3,\,2/3,\,1/3$, fixed $c_{\rm s}^2=0.1$, $w=-0.9$ and 
the 
same $\kappa$ and $C$. (iv) To analyse the dependence of void parameters 
on its scale we run the code for $k\equiv\kappa/\pi=0.001,\,0.05,\,0.1,\,0.2$ 
Mpc$^{-1}$ and fixed $w=-0.9$,
$c_{\rm s}^2=0.1$, $\kappa^2r_{\rm d}^2=4/3$, $C=-2\times10^{-4}$. (v) To 
analyse the dependence of void parameters on amplitude of initial
perturbation we run the code for 
$C=-0.5\times10^{-4},\,-1\times10^{-4},\,-2\times10^{-4}$ and fixed $w=-0.9$, 
$c_{\rm s}^2=0.1$,
$\kappa^{-1}=62.8$ Mpc and $\kappa^2r_{\rm d}^2=4/3$. The results of two runs 
from (i) are presented in Figs \ref{cs00} and \ref{cs01},
the results of three runs from (iii) are presented in Fig. \ref{profiles}. The 
discussion of the results of all runs are subject 
of the next sections.  


\section{Formation of voids in the cosmological models with dark energy}

The results of integration of the system of evolution equations 
(\ref{m_cl0})-(\ref{ee00_l2}) for initial perturbation, which is shown in the 
left-hand column of Fig. \ref{init}, are shown in Fig. \ref{cs00} for the model 
with dark energy parameters $\Omega_{\rm de}=0.7$, $w=-0.9$ and $c_{\rm 
s}^2=0$. 
The black thick solid lines denote the initial profiles of density and velocity 
perturbations of matter (left-hand column) and dark energy (central column) at 
$a_{\rm init}=10^{-6}$, the red thick dash--dotted lines 
denote the final ones at $a=1$, the blue thin solid ones show for intermediate 
values of $a$.  The figure on the right depicts the evolution of absolute 
values 
of amplitudes of perturbations in the central point of spherical void and in 
the 
overdense
shell. Velocity perturbation (bottom panel) are given for the first maximum 
(circles and squares, which superimposed) and the first
minimum (triangles and fourfold stars, which superimposed). The signs in the 
right-hand column correspond to the values in the mentioned
positions in the corresponding curves in the left-hand and central columns. 
Dotted lines denote the evolution of the same values for the
radiation component. One can see that in this dark energy model the 
perturbations of matter and dark energy grow monotonically
after entering the horizon: the black lines are internal, the red lines are 
external. We also note that the amplitude of the
density perturbation of dark energy is approximately 40 times smaller than the 
one of matter. The values of velocity perturbations
of matter and dark energy in this model of dark energy are the same throughout 
the evolution of the void. They increase
monotonically from $a_{\rm init}$ to $a\approx0.56$. It is easy to see that the 
latter value corresponds to the moment of change from
the decelerated expansion of the Universe to the accelerated one. The evolution 
of the absolute values of density and velocity
perturbations of matter and dark energy in the overdense shell is similar to 
the 
evolution of those in centre. 

The similar results of modelling of the void formation in the matter and dark 
energy with $c_{\rm s}^2=0.1$ are shown in Fig. \ref{cs01}. One can
see that evolution of matter density and velocity perturbations is practically 
the same, while for dark energy it has changed drastically.
The final profiles of dark energy perturbations (red dash--dotted lines) have 
very small amplitudes. It means that dark energy in the voids is
only slightly perturbed. The right figure explains such behaviour of dark 
energy 
during the void formation: the velocity perturbation after
the entering into horizon decrease quickly, and density perturbation slightly 
changes during all stages and in the current epoch $\rho_{\rm de}(1,0)$ does not
differ significantly from the background value: $\delta_{\rm 
de}(1,0)\approx-9\times10^{-6}$. We see also that the evolution of the absolute 
values
of density and velocity perturbations of dark energy in the overdense shell 
slightly differ from the evolution of ones in the centre of the
void. At the current epoch they are approximately equal small magnitudes. 
 
The perturbations of dark energy with larger values of effective speed of sound 
after entering the particle horizon is smoothed out even faster. The ratio of 
densities of dark energy and matter in the centre of the void is
$$\frac{\rho_{\rm de}(1,0)}{\rho_{\rm m}(1,0)}=\frac{1+\delta_{\rm 
de}(1,0)}{1+\delta_{\rm m}(1,0)}\frac{\Omega_{\rm de}}{\Omega_{\rm m}},$$
and in the case of evolution with considered initial condition this ratio is 
two 
to five times larger than on cosmological background.  
\begin{table*}
\centering
 \normalsize
 \caption{The final parameters of voids (matter density perturbations at the 
centre $\delta_{\rm c}$, matter density perturbation in the overdense shell 
$\delta_{\rm sh}$, distance $r_{\rm sh}$ where shell is densest, maximal 
peculiar velocity of matter $\mathpzc{v}_{\rm m}^{\rm (max)}$, distance $r_{\rm 
mv}$ where $\mathpzc{v}_{\rm m}=\mathpzc{v}_{\rm m}^{\rm (max)}$, final size of 
void $r_{\delta=0}^{\rm (final)}$) with different initial parameters 
$r_{\delta=0}^{\rm (init)}$, $r_{\rm d}$ and $C$. The rest of parameters are 
the 
same as in Fig. \ref{cs01}.}
\begin{tabular}{c|c|c|c|c|c|c|c|c|c}
 \hline  
&&&&&&&&&\\ 
$r_{\delta=0}^{\rm (init)}$,&$r_{\rm d}$,&$C\times10^4$&$\delta_{\rm 
c}$&$\delta_{\rm sh}$&$r_{\rm sh}$,&$\mathpzc{v}_{\rm m}^{\rm (max)}$,&$r_{\rm 
mv}$,& $\frac{\mathpzc{v}_{\rm m}^{\rm (max)}}{\mathpzc{v}_{\rm 
H}}$&$r_{\delta=0}^{\rm (final)}$,\\
(Mpc)&(Mpc)&&&&(Mpc)&(km s$^{-1}$)&(Mpc)&&(Mpc)\\
 \hline  
31.4&36.3&-1&-0.68&0.33&45.7&202&29.9&0.097&36.9\\   
31.4&25.6&-1&-0.69&0.16&38.6&169&24.6&0.098&29.9\\  
31.4&18.1&-1&-0.73&0.09&31.8&148&20.1&0.11&25.1\\
62.8&72.5&-1&-0.51&0.14&91.3&237&52.7&0.064&68.5\\
62.8&51.3&-1&-0.54&0.09&73.8&204&40.4&0.072&57.9\\
62.8&36.3&-1&-0.59&0.06&59.7&186&33.4&0.080&47.4\\
62.8&72.5&-2&-0.70&0.34&91.3&428&59.7&0.10&73.8\\
62.8&51.3&-2&-0.72&0.20&73.8&365&47.4&0.11&61.5\\
62.8&36.3&-2&-0.77&0.14&59.7&329&40.4&0.12&50.9\\
\hline
\label{final_void_par}
\end{tabular}
\end{table*} 

The matter density perturbation in the central part of this void at the current 
epoch is $\delta_{\rm c}\approx-0.7$, the magnitude of density 
perturbation in the overdense shell is $\delta_{\rm sh}\approx0.34$, these 
values are close to the mean ones in the real voids \citep{Mao2016}.  The 
maximum of peculiar velocity of matter which move from centre is 
$\mathpzc{v}_{\rm m}^{\rm (max)}\approx1.43\times10^{-3}c\approx428$ km 
s$^{-1}$ 
at $r_{\rm mv}\approx59.7$ Mpc. The Hubble flow at such distance from the 
centre 
is $V_{\rm H}=H_0r_{\rm mv}\approx4179$ km s$^{-1}$, so the maximum 
of 
peculiar velocity of matter in units of Hubble one at the same distance is 
$\mathpzc{v}_{\rm m}^{\rm (max)}/V_{\rm H}\approx0.1$. The distance 
where the matter density perturbation becomes zero is now $r_{\delta=0}^{\rm 
(final)}\approx73.8$ Mpc. 
Thus, in this model the comoving radius of void has increased approximately in 
1.2 times. The same final parameters of voids for different initial 
$r_{\delta=0}^{\rm (init)}$, $r_{\rm d}$ and $C$ are presented in the Table 
\ref{final_void_par}. It gives us the possibility to understand 
how each of parameters of initial profile affects the parameters of final 
voids. 
For example, increasing the size of seed perturbation $r_{\delta=0}^{\rm 
(init)}$ 
with unchanged other parameters (first and sixth rows) results in increased 
final size $r_{\delta=0}^{\rm (final)}$ but decreased magnitude of density 
perturbations in the centre $\delta_{\rm c}$ and in the overdense shell 
$\delta_{\rm sh}$ as well as the value of 
maximal peculiar velocity $\mathpzc{v}_{\rm m}^{\rm (max)}$. The decreasing of 
$r_{\rm d}$ with unchanged other parameters (rows 1-3 rows 4-6, rows 7-9) leads 
to
decreasing of void size, maximal value of peculiar velocity, magnitude of 
density perturbations in the overdense shell and matter density at
the void centre. If $r_{\rm d}<r_{\delta=0}^{\rm (init)}$ then the final void 
size 
$r_{\delta=0}^{\rm (final)}$ becomes smaller than initial one (third, sixth
and ninth rows). The decreasing of $\kappa$ (increasing of $r_{\delta=0}$) for 
fixed $\kappa^2r_{\rm d}^2$ and the same rest parameters
leads to the decrease of final amplitude of density perturbation and to the 
increase of final amplitude of peculiar velocity in the void as well as in the 
overdensity shell. However, comparing the final values of void parameters in 
rows 1--4, 2--5 and 3--6 of the Table \ref{final_void_par} shows that they are 
not simply rescaled, as it might seem by visual comparison of Figs. \ref{cs00} 
and \ref{cs01} of this paper with rigs. 1 and 2 in our paper \citep{Tsizh16}. 
The increasing of initial amplitude $C$ with unchanged other parameters 
(comparing of values in the rows 4, 5, 6 with corresponding values in the rows 
7, 8, 9) leads to increasing the final size of void, the amplitude of
density perturbations at the centre and in overdense shell as well as the value 
of maximal velocity of matter.  
\begin{figure*}
\includegraphics[width=0.33\textwidth]{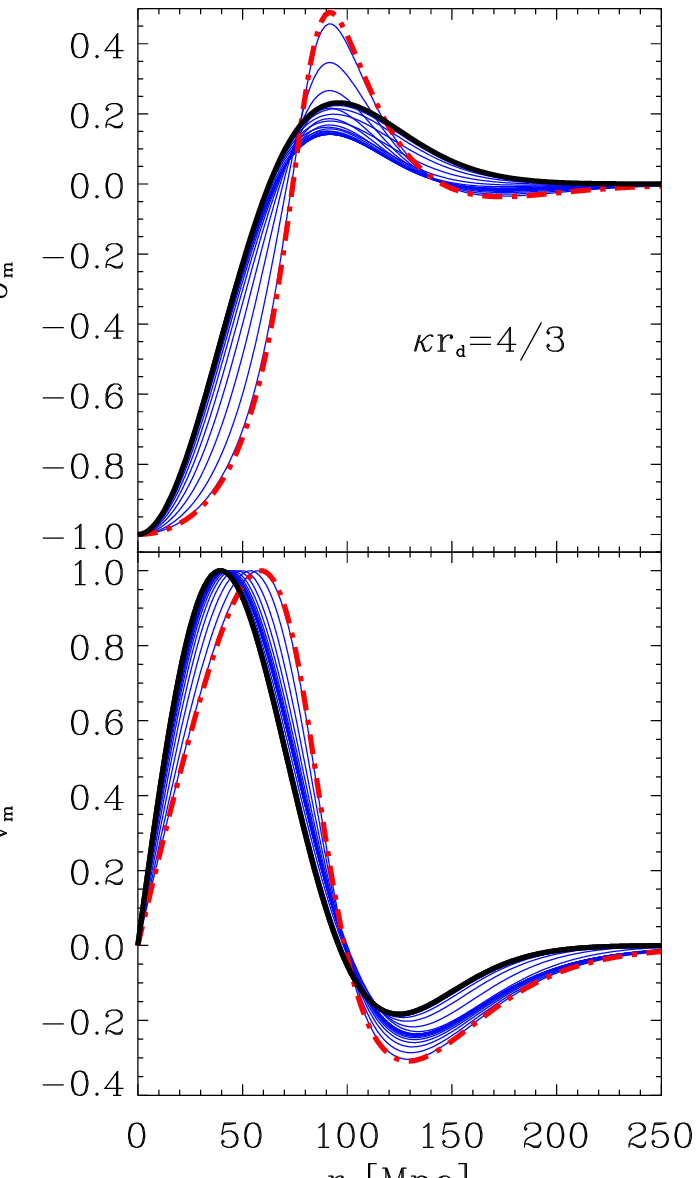} 
\includegraphics[width=0.33\textwidth]{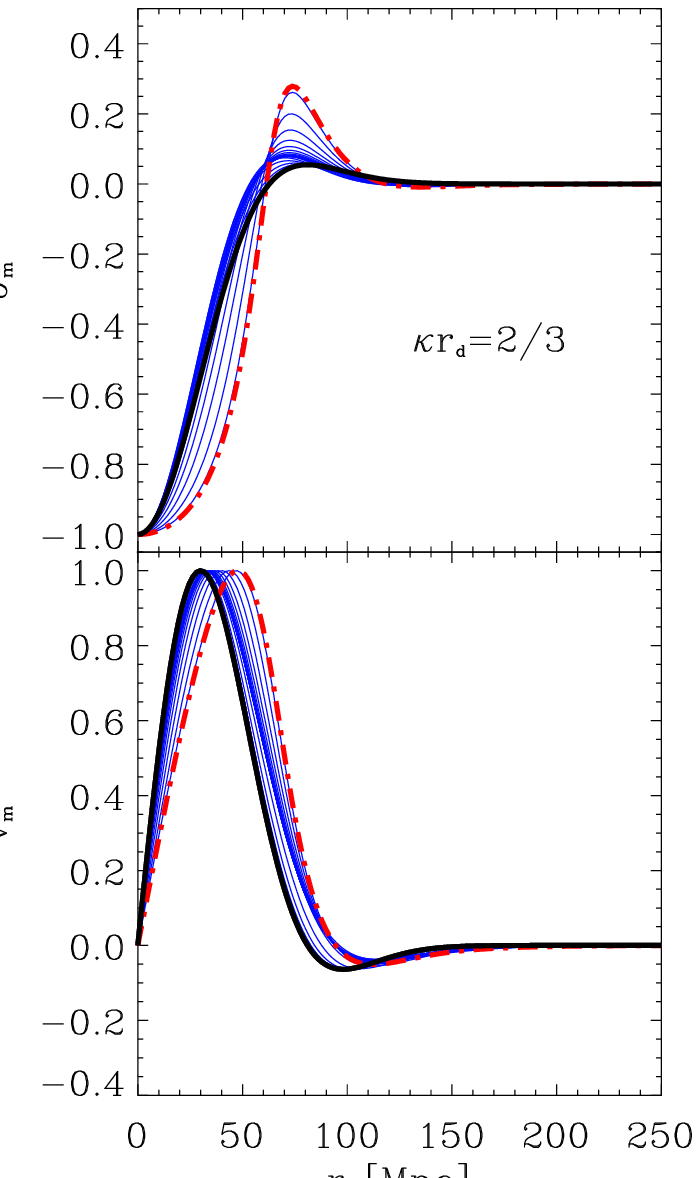}
\includegraphics[width=0.33\textwidth]{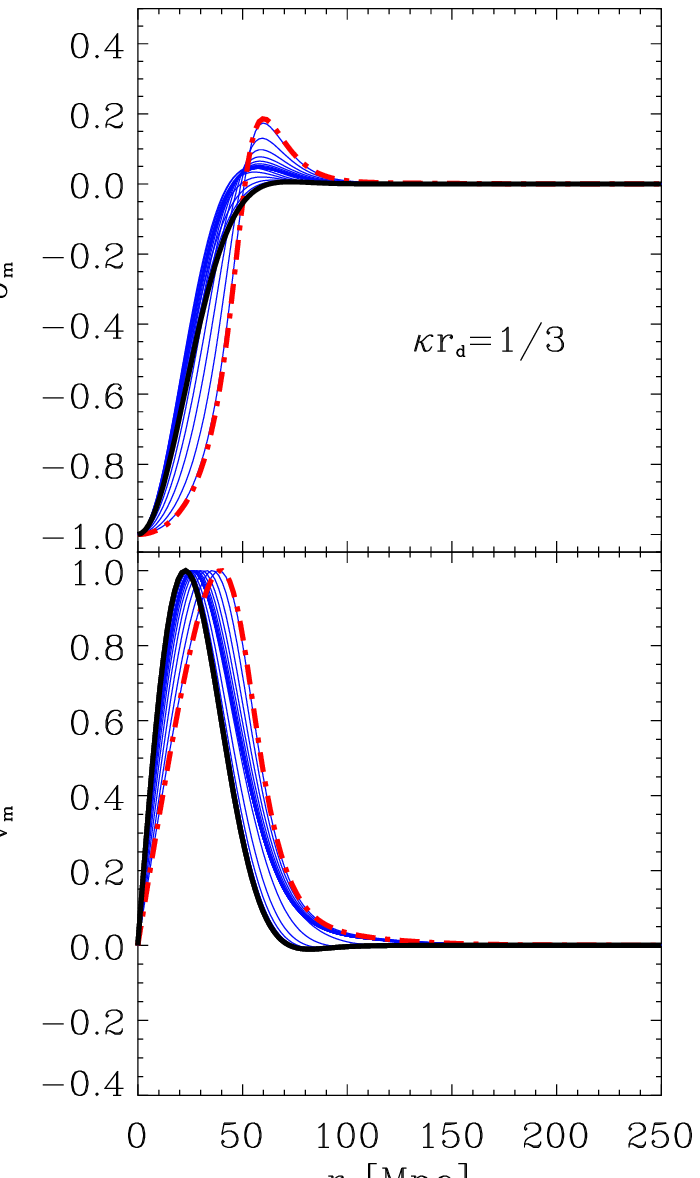}  
\caption{Evolution of normalized profiles of matter density and velocity 
perturbations with initial parameters $\kappa=(62.8)^{-1}$ Mpc$^{-1}$ and 
$\kappa^2r_{\rm d}^2=4/3$ (left-hand), 2/3 (central) and 1/3 (right-hand 
column). Thick black solid lines are normalized initial profiles (Fig. 
\ref{init}), thick red dash--dotted ones are final profiles and blue lines are 
for intermediate increasing values of $a$. Other parameters are the same as in 
Fig. \ref{cs01}.}
\label{profiles}
\end{figure*}

Let us analyse the evolution of matter density profile regardless of their 
amplitudes. For that, we renormalize each curve in the left-hand columns of 
Figs 
\ref{cs00} and \ref{cs01} by its amplitude. The results are shown in Fig. 
\ref{profiles} for initial perturbations with three different values of 
$\kappa^2r_{\rm d}^2=4/3,\,2/3,\,1/3$ and the same other parameters as in 
previous figures. First of all, we note that the profiles of
matter density and velocity perturbations vary with time. They become less 
steep 
at the centre and more steep at the edge of void and in the region of overdense 
shell. One can also see that overdense shell appears in the process of 
evolution of void even if its amplitude was very small in the initial profile 
(figure on the right), or absence at all ($\kappa=0$, Gaussian initial 
profile). 
The relations between final amplitudes of shell density perturbations have 
changed essentially: $\delta_{\rm sh}^{(4/3)}:\delta_{\rm 
sh}^{(2/3)}:\delta_{\rm sh}^{(1/3)}\approx 2.4:1.5:1$ for voids with 
$r_{\delta=0}^{\rm (init)}=62.8$ Mpc and $C=-2\times10^{-4}$. In the units of 
magnitudes of central density perturbations the magnitudes of overdense shells 
for these profiles now are 0.49, 0.28 and 0.19 against 0.23, 0.055 and 0.006 
for 
initial profiles. 
They depend also on initial amplitude and scale of the void seeds of the same 
profile. For example, for voids with $r_{\delta=0}^{\rm (init)}=62.8$ Mpc and 
$C=-1\times10^{-4}$ the final magnitudes of overdense shells in the units of 
magnitudes of central density perturbations are 0.28, 0.16 and 0.10 for three 
profiles accordingly. For voids with $r_{\delta=0}^{\rm (init)}=31.4$ Mpc and 
the 
same $C$ these numbers are 0.49, 0.23 and 0.13, close to the ones for voids 
with 
$r_{\delta=0}^{\rm (init)}=62.8$ Mpc and $C=-2\times10^{-4}$. It explains the 
similarity of normalized profiles in Fig. \ref{profiles} and in fig. 3 of 
\cite{Tsizh16}.

We must to note also that voids with considered sizes and amplitudes (see Table 
\ref{final_void_par}) do not show tendency to collapse of voids with an 
overdense shell, which have been discussed in \cite{Sheth2004} and 
\cite{Ceccarelli13}. The void size $r_{\delta=0}$ increases monotonically with 
$a$ from $a_{\rm init}$ up to $a=1$ in the left-hand panel of Fig. 
\ref{profiles}, and increases monotonically after formation of overdense shell 
in the central and right-hand ones. Similar behaviour is demonstrated also by 
position of the peak density of overdense shells 
$r_{\rm sh}\equiv r_{max\,\{\delta_{\rm m}\}}$. For the final profiles in the 
top panels of Fig. \ref{profiles} they are $r_{\rm sh}=91.3,\,73.8,\,59.7$ Mpc 
from left to right accordingly.
It can be associated with watershed ridgelines in the algorithms of void 
finding 
in the galaxies catalogues and snapshots of $N$-body simulations    
\citep{Platen07,Neyrinck08,Sutter14a}. One can consider also the radius of 
overdense shell as the radius of the sphere on which the peculiar velocity of 
matter is zero $r_{\mathpzc{v}_{\rm m}=0}$ and change of sign from `+' to `--'. 
For the final profiles in the bottom left-hand and central panels of Fig. 
\ref{profiles} they are $r_{\mathpzc{v}_{\rm m}=0}=98.3,\,94.8$ Mpc, 
respectively. The peculiar velocity shown in the bottom-right panels is 
positive 
in all range of $r$. The value of $r_{\mathpzc{v}_{\rm m}=0}$ grows with time 
too, as it is shown in the figures. So, since these scale parameters are in the 
frame, which is comoving to cosmological background, the stage of evolution is 
far from turnaround point, which is begin of the collapse.   

\begin{figure*}
\begin{center}
\includegraphics[width=0.45\textwidth]{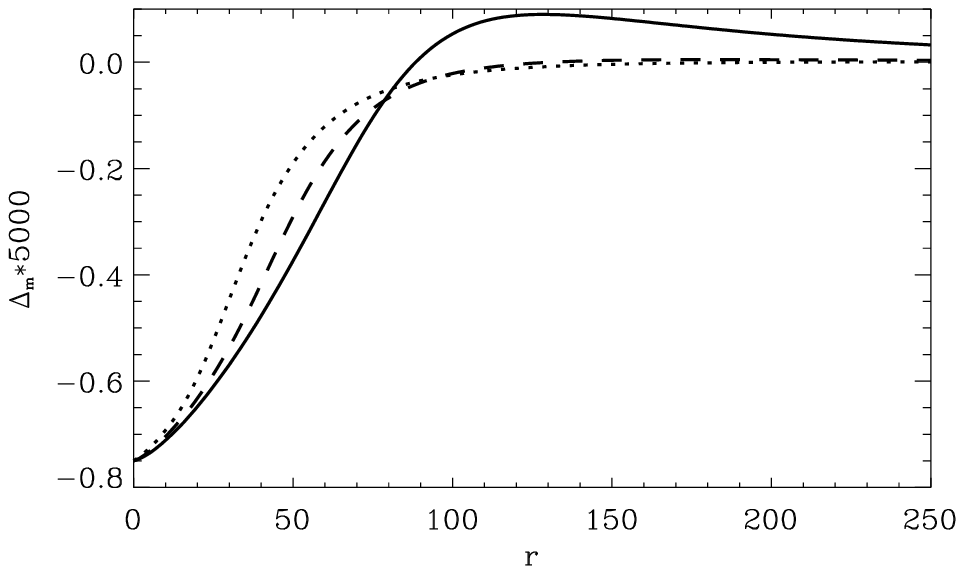} 
\includegraphics[width=0.45\textwidth]{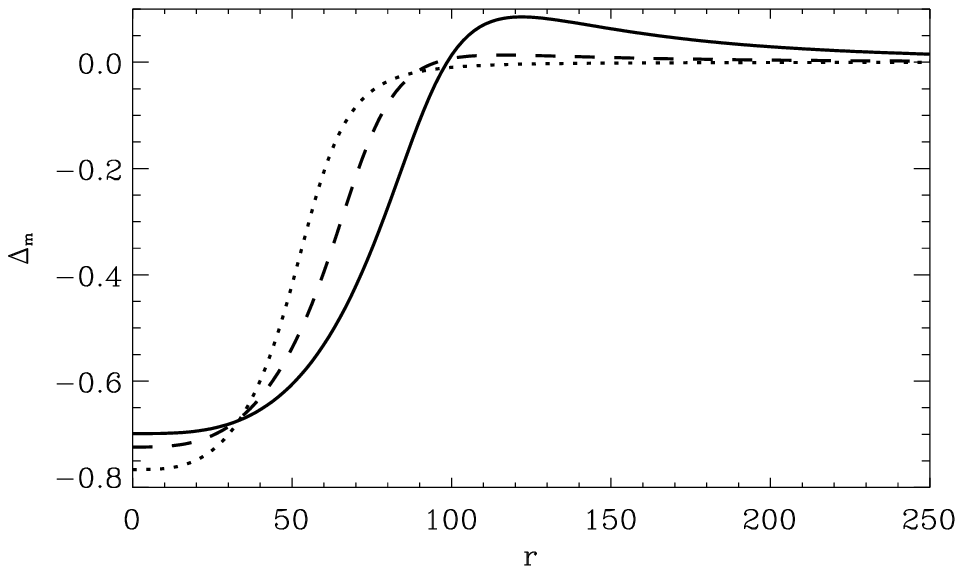}
\end{center}   
\caption{The relative perturbations of mass of matter for three profiles with 
$\kappa^2r_{\rm d}^2=4/3$ (solid lines), 2/3 (dashed lines) and 1/3 (dotted 
lines) for initial matter density perturbations profiles (left-hand panel) and 
for final ones (right-hand panel).}
\label{fig_dm}
\end{figure*}
Now consider how the relative perturbation of mass of matter in the sphere of 
radius $r$, which equals the integrated density contrast,        
\begin{equation}
\Delta_{\rm m}(r)=\frac{3}{r^3}\int_0^r\delta_{\rm m}(q)q^2dq, \label{dm}
\end{equation} 
is changed under void formation. In the left-hand panel of Fig. \ref{fig_dm}, 
we 
present the relative perturbations of mass of matter for three initial profiles 
of matter density perturbations and in the right-hand panel for the final ones. 
One can see that relative perturbation of mass for density profile with 
$\kappa^2r_{\rm d}^2=4/3$ change the sign in the region of overdense shell at 
the distance $r_{\Delta=0}=98.3$ Mpc and for outer observer it is positive 
perturbations of mass (overcompensated void or void-in-cloud; 
\citep{Sheth2004}). Such voids may exhibit infall velocities of matter in the 
outer layers of overdense shells and neighbour voids may gravitationally 
attract.
For density profile with $\kappa^2r_{\rm d}^2=2/3$ the relative perturbation of 
mass becomes zero inside the shell at the distance $r_{\Delta=0}=94.8$ Mpc and 
remains so for larger distances from the centre. Note also that 
$r_{\Delta=0}=r_{\mathpzc{v}_{\rm m}=0}$, which was expected. For density 
profile with $\kappa^2r_{\rm d}^2=1/3$ the relative perturbation of mass does 
not change the sign 
and remains so for larger distances from the centre and for outer observer it 
is 
negative perturbations of mass (undercompensated void). Such neighbour voids 
may 
coalesce and form larger non-spherical void. The result of evolution of void is 
increasing of amplitude of mass perturbation almost 5000 times and 
redistribution of matter between the marginal part of void and overdense shell 
(see profiles for peculiar velocity in Figs.\ref{cs00}-\ref{profiles}). The 
possibility of collapse of voids with overdense shell is determined by the sign 
of total mechanical energy of the system, not only by the sign of perturbation 
of total mass. 

\section{Universal void profile and its evolution} 

Let us compare our profiles with universal density profiles for cosmic voids 
proposed by \cite{Hamaus14}:
\begin{equation}
\delta_{\rm m}(r)=\delta_{\rm c}\left[1-\left(\frac{r}{r_{\rm 
s}}\right)^\alpha\right]/\left[1+\left(\frac{r}{r_{\rm 
\mathpzc{v}}}\right)^\beta\right].\label{udp}
\end{equation}
We approximate the final matter density profile matching parameters $r_{\rm 
s}$, 
$\alpha$, $r_{\rm \mathpzc{v}}$ and $\beta$  by the Levenberg--Marquardt method 
\citep{NumRec1993}. For that, we have used the subroutine \textit{mrqmin.f} 
from 
the Numerical Recipes Library. The value of $\delta_{\rm c}$ we take equal  
$\delta_{\rm m}(a,0)$ in each specific profile.  
In the left-hand panel of Fig. 6 we show the matter density perturbation 
profile 
for the void model with parameters from the last row of Table 
\ref{final_void_par} and its approximation by equation (\ref{udp}) with 
best-fitting parameters $r_{\rm s}=51,457$ Mpc, $\alpha=3.2141$, $r_{\rm 
\mathpzc{v}}=55.709$ and $\beta=11.31$. These lines are superimposed on one 
another. 

In the right-hand panel of Fig. 6 we show the matter peculiar velocity profile 
for the same void model. The dotted line presents the peculiar velocity profile 
which follows from the linear theory of cosmological perturbations  
$\mathpzc{v}_{\rm lin}(r)\approx-\frac{1}{3}\Omega_{\rm m}^{\gamma}\Delta_{\rm 
m}(r)H_0r$, where $\Delta_{\rm m}(r)$ is relative perturbation of mass 
(\ref{dm}) and $\gamma\approx0.55$ \citep{Peebles80}. It coincides with 
velocity 
profile computed by formula 6 from \cite{Hamaus14} with parameters   
presented above for density. In the range of maximum of velocity they differ by 
factor of $\sim1.2$, since the later stages of void 
evolution are not linear. That is why we propose other simple formula for 
approximation of peculiar velocity profile of matter in the spherical voids:
\begin{equation}
\mathrm{v}_{\rm m}(r)=10^{-3}\left(\frac{r}{r_{\rm lv}}+\frac{r^3}{r_{\rm 
3v}^3}-\frac{r^5}{r_{\rm 5v}^3}\right)e^{-r^2/r_{\rm dv}^2}\,\,\,{\rm km \, 
s^{-1}},\label{ccpg}
\end{equation}
which is the convolution of odd degree polynomial and Gaussian. In the Fig. 6, 
this approximation with best-fitting parameters $r_{\rm lv}=30.95$ Mpc, $r_{\rm 
3v}=28.73$ Mpc, $r_{\rm 5v}=49.05$ Mpc and $r_{\rm dv}=36.33$ Mpc is shown by 
solid 
thin blue line. For the $r_{\rm mv}=40.4$ Mpc, we obtain $\mathrm{v}_{\rm 
m}^{\rm max}\approx330$ km s$^{-1}$. The velocity computed according to the 
formula of linear theory in this point is $\mathpzc{v}_{\rm lin}^{\rm 
max}\approx290$ km s$^{-1}$. This approximation formula for peculiar velocity 
has the property that in the centre part of the void the velocity is 
proportional to the distance.  

\begin{figure*}
\begin{center}
\includegraphics[width=0.9\textwidth]{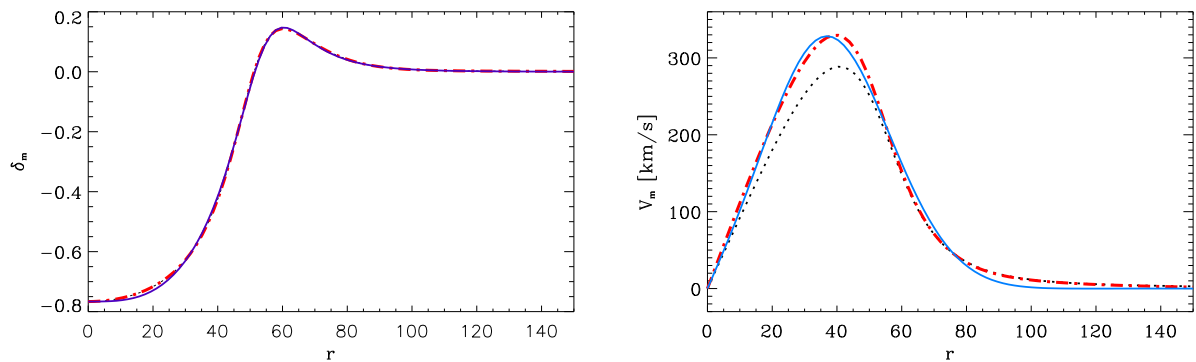}  
\caption{Left-hand panel: the final profile of matter density for the model 
with parameters from the last row of Table \ref{final_void_par} (dash--dotted 
red 
line) and its approximation (solid thin blue line) by universal density profile 
(equation \ref{udp}). Right-hand panel: the final profile of matter peculiar 
velocity (dash--dotted red line) and its approximation (solid thin blue line) 
by 
convolution of cubic parabola and Gaussian (equation \ref{ccpg}). The dotted 
line is linear velocity $\mathpzc{v}_{\rm lin}$ for density profile from the 
left-hand panel.}
\end{center}
\label{appr}
\end{figure*}

\begin{figure*}
\includegraphics[width=0.48\textwidth]{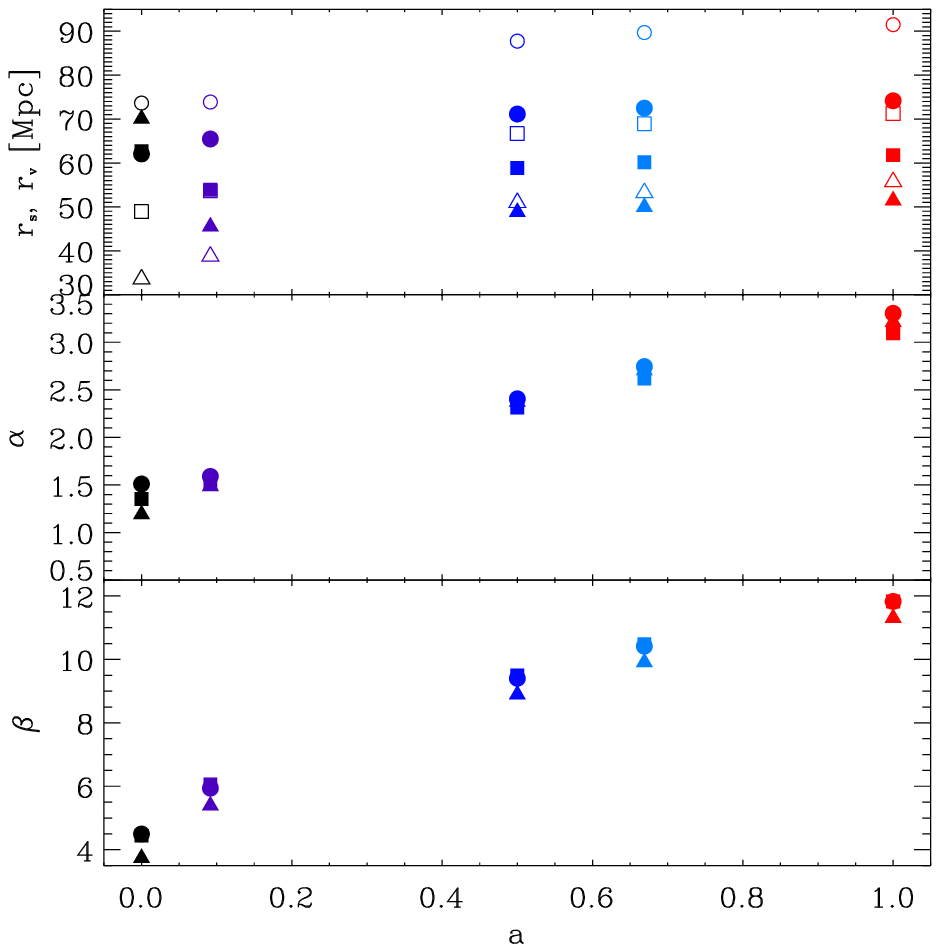} 
\includegraphics[width=0.48\textwidth]{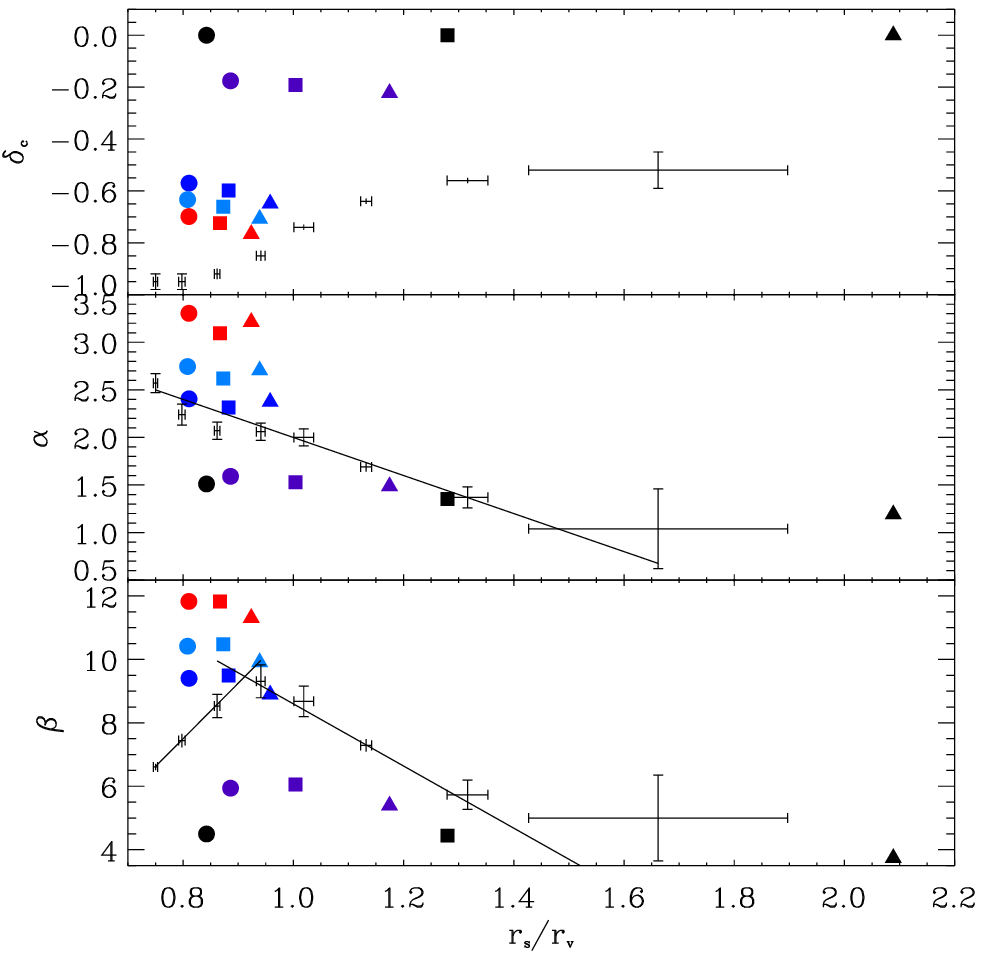}  
\caption{Left-hand panels: the evolution of the best-fitting values of 
parameters of universal void profile ($r_{\rm s}$, $r_{\rm \mathpzc{v}}$, 
$\alpha$ and $\beta$) matching the matter density perturbations at different 
redshifts (from left to right: black at $z=10^6$, dark blue at $z=10$, blue at 
$z=1$, light blue at $z=0.5$ and red at $z=0$) and different initial profile 
(cirques for $\kappa^2r_{\rm d}^2=4/3$, squares for $\kappa^2r_{\rm d}^2=2/3$ 
and triangles for $\kappa^2r_{\rm d}^2=1/3$). In the top left-hand panel the 
open symbols are for $r_{\rm \mathpzc{v}}$, closed for $r_{\rm s}$. Right-hand 
panels: comparison of the best-fitting values of parameters of universal void 
profile for our voids with parameters of profiles of voids obtained from 
$N$-body simulations by Hamaus et al. (2014) (points with error bars and 
lines). 
In the top panel the evolution goes top-down and in the middle and bottom ones 
it goes bottom-top. The initial radius of all voids here is 62.8 Mpc, the 
initial 
amplitude is $C=-2\times10^{-4}$.}
\label{uv_par}
\end{figure*}

Let us look how parameters $r_{\rm s}$, $\alpha$, $r_{\rm \mathpzc{v}}$ and 
$\beta$ changed for time of evolution. For that, we approximate the matter 
density profiles with $\kappa^2r_{\rm d}^2=4/3,\,\,2/3$ and 1/3 at redshifts 
$z=10^6,\,10,\,1,\,0.5$ and 0 matching the parameters of the universal density 
profile by the Levenberg--Marquardt method for each profile computed in our 
models 
of voids. The results are presented in the Fig. \ref{uv_par} by different 
symbols (cirque, square, triangle) for different initial profiles 
($\kappa^2r_{\rm d}^2=4/3$, 2/3, 1/3 accordingly) and by different colours for 
different redshifts (black for $z=10^6$, dark blue for $z=10$, blue for $z=1$, 
light blue for $z=0.5$ and red for $z=0$). The evolution of parameters is shown 
in the left-hand panel of Fig. \ref{uv_par}. One can see that scale parameters 
$r_{\rm s}$ and $r_{\rm \mathpzc{v}}$  as well as power-law  parameters 
$\alpha$ 
and $\beta$ monotonically increase, which reflect the increase of sizes and 
amplitude of voids and surrounding overdense shells as well as steepness of 
their walls.

To compare our profiles of spherical voids with stacked ones from $N$-body 
simulations we impose these parameters of universal profiles on fig. 2 of 
\cite{Hamaus14}. In the right-hand panel of Fig. \ref{uv_par} we show the 
parameters $\delta_{\rm c}$, $\alpha$ and $\beta$ for corresponding $r_{\rm 
s}/r_{\rm \mathpzc{v}}$. One can see that parameters of profiles of our voids 
at 
different redshifts occupy mainly the same regions of parameter space where 
voids from $N$-body simulations are stacked. 

We note also that void profiles, obtained here by modelling of their formation 
from cosmological perturbations with spherical initial 
profiles, have parameters close to stacked voids from the galaxy catalogues 
\citep{Sutter12,Nadathur14,Mao2016}. 

\section{Sensitivity of void parameters to dark energy ones}

We estimated how amplitudes of density and velocity perturbations in the void 
models are sensitive to the dark energy parameters,
especially to $c_{\rm s}^2$ and $w$. For that we have computed the evolution of 
perturbations with the same initial conditions and different
$c_{\rm s}^2=0,\,0.01,\,0.1,\,1$ and $w=-0.8,\,-0.9,\,-1,\,-1.1,\,-1.2$. The 
results show that the amplitudes of density perturbations of the void 
$\delta_{\rm c}$ with parameters in Table \ref{final_void_par} for $c_{\rm 
s}^2=0$ and 1 differ in $\le0.2$ per cent, the amplitudes of density 
perturbations of the overdense shell $\delta_{\rm sh}$ differ in $\le0.5$ per 
cent and the maximal values of matter peculiar velocity $\mathpzc{v}_{\rm 
m}^{\rm max}$ differ in $\le1.6$  per cent. 
So, the parameters of spherical void models very slightly depend on effective 
sound speed of dark energy. But the same parameters of spherical void models 
depend strongly on the value of equation-of-state parameter $w$. Comparing 
$\delta_{\rm c}$, $\delta_{\rm sh}$ and $\mathpzc{v}_{\rm m}^{\rm max}$ for 
$w=-0.8$ and --1 we obtain the difference in $\sim4$ per cent, $\sim9$ per cent 
and $\sim7$ per cent 
respectively. Comparison of $\delta_{\rm c}$, $\delta_{\rm sh}$ and 
$\mathpzc{v}_{\rm m}^{\rm max}$ for $w=-1.2$ and -1 gives differences in 
$\sim2$ per cent, $\sim6$ per cent and $\sim4$ per cent, respectively, but with 
opposite sign.  
Therefore, an accurate measurement of matter density and peculiar velocities of 
galaxies in the voids can be powerful discriminator between quintessence, 
phantom and $\Lambda$ models of dark energy. 

\section{Conclusions}

The large voids in the spatial distribution of galaxies are formed from the 
large-scale dips in the Gaussian random field of primordial  density 
perturbations. Such dips with initial radius $r_{\rm s}\sim30-60$ Mpc (in 
comoving coordinates) and initial amplitudes $\sim-1\times10^{-4}$ to 
$-2\times10^{-4}$ 
($\sim$1--3 rms value in the concordance $\Lambda$CDM model) result into voids 
with central density perturbation $\sim-0.5$ to $-0.8$ and maximal value of 
peculiar velocity $\sim$ 150--430 km s$^{-1}$ which is directed outward and is 
reached at the distance $\sim$0.7--0.8$r_{\rm s}$ from the centre of the void. 
The overdense shell is formed around void by  raking of outflowing matter 
and/or 
infall outer layers. Its position and amplitude of density depend on the 
amplitude and type of initial profile, overcompensated or undercompensated. The 
profile of density perturbations of void evolves in such a way that comoving 
scale parameters of universal profile \citep{Hamaus14} $r_{\rm s}$ and $r_{\rm 
\mathpzc{v}}$ are increased by $\sim$1.1--1.2 times, the 
steepness parameters $\alpha$ and $\beta$ are increased by $\sim$2.2--3.0 
times. 

The density and velocity perturbations of the dark energy evolve similarly to 
the perturbations of matter at the stage when their scales are much larger than 
the particle horizon. After they enter the particle horizon their evolution 
depends on the value of the effective speed of sound $c_{\rm s}$. If $c_{\rm 
s}=0$, then similarity is conserved with the difference that the amplitude of 
density perturbation of dark energy is smaller in factor $1+w$. At the later 
epoch, when the dark energy density dominates, this difference increased yet in 
$\approx$4--5 times more. If $0<c_{\rm s}\le1$, then the amplitude of velocity 
perturbation of dark energy after entering the horizon decreases rapidly, the 
amplitude of the density perturbation does not increase or even decreases too. 
Therefore, in the voids the density of dynamical dark energy is approximately 
the same as in the  cosmological background. The ratio of the densities of dark 
energy and matter is  $1/(1+\delta_{\rm m})$ larger than that in the 
cosmological background. The more hollow void is the larger this ratio is. Our 
estimations show, that measurable parameters of large spherical voids with size 
$\sim60$ Mpc and typical amplitude of initial perturbation are sensitive to 
equation-of-state parameters of dark energy $w$ at the level of few per cents 
and to effective sound speed at the level of 1 per cent. It means that 
investigation of structure and dynamics of large voids are perspective for 
testing of models of dark energy and gravity modifications.

\section*{Acknowledgements}
This work was supported by the project of Ministry of Education and Science of 
Ukraine with state registration number 0116U001544.  

\label{lastpage}
\end{document}